\documentclass{pasj00}

\SetRunningHead{S.Torii et al.}{Spectral and Timing Studies of Cyg X-1 in the LHS with Suzaku}

\draft
\usepackage{color}

\usepackage{times}

\usepackage{multicol}

\usepackage{natbib}

\usepackage{ulem}

\begin{document}

\title{Spectral and Timing Studies of Cyg X-1 in the Low/Hard State with Suzaku}
\author{Shunsuke TORII$^1$, Shin'ya YAMADA$^2$, Kazuo MAKISHIMA$^1$, Soki SAKURAI$^1$,\\
Kazuhiro NAKAZAWA$^1$, Hirofumi NODA$^1$, Chris DONE$^3$, Hiromitsu TAKAHASHI$^4$\\
and Poshak GANDHI$^5$}
\affil{$^1$Department of Physics, The University of Tokyo, 7-3-1 Hongo, Bunkyo-ku, Tokyo 113-0033}
\email{torii@juno.phys.s.u-tokyo.ac.jp}
\affil{$^2$Cosmic Radiation Laboratory, Institute of Physical and Chemical Research (RIKEN), 2-1 Hirosawa, Wako-shi, Saitama 351-0198}
\affil{$^3$Department of Physics, Durham University, South Road, Durham, DH1 3LE, UK}
\affil{$^4$Department of Physical Science, School of Science, Hiroshima University, 1-3-1 Kagamiyama, Higashi-Hiroshima, Hiroshima 739-8526}
\affil{$^5$Institute of Space and Astronautical Science, JAXA, 3-1-1 Yoshinodai, Sagamihara, Kanagawa 229-8510}

\maketitle

\begin{abstract}

From 2005 to 2009, 25 observations of Cyg X-1 were
performed with Suzaku, achieving a total exposure of 446 ks.
In all observations, the source was found in the low/hard state,
while the 1.5--12.0 keV count rate of the All-Sky Monitor onboard RXTE varied by a factor of $\sim$ 3.
In each observation, the 10--60 keV HXD-PIN spectrum and the 60--400
keV HXD-GSO spectrum were  fitted successfully by a thermal Comptonization model plus reflection
by a thick neutral material.
As the soft X-ray intensity increased,
the Compton $y$-parameter was found to decrease from 1.0 to 0.6,
while the solid angle of reflection to increase by $\sim 30\%$.
Also conducted was timing analysis over a frequency range of 10$^{-3}$--10 Hz.
As the source became brighter in soft X-rays,
the characteristic frequency of hard X-ray variation
increased from 0.03 to 0.3 Hz,
while the fractional hard X-ray variation integrated
over 10$^{-3}$--10$^{-2}$  Hz decreased by a factor of $\sim$ 5.
The signals in the 60--200 keV band were generally found to
vary on shorter time scales than those in the 10--60 keV band.
These spectral and timing results can be consistently
interpreted by presuming that increases in the mass accretion rate
cause the Comptonizing hot corona to shrink,
while the optically-thick disk to intrude deeper therein.

\end{abstract}
\section{Introduction}

Black hole binaries (BHBs) are known to reside mainly in either of two spectral states, the low/hard state (LHS) and the high/soft state (HSS),
which appear when mass accretion rate is lower and higher, respectively (McClintock \& Remillard 2006).
In the HSS, the accreting matter is considered to release its energy predominantly in soft X-rays as a multi-temperature disk emission
(Mitsuda et al. 1984, Makishima et al, 1986) from an optically thick and geometrically thin accretion disk, called standard disk (Shakura \& Sunyaev 1973).
In the LHS, in contrast, the spectrum becomes much harder, and is characterized by a flat power-law with a photon index of $\Gamma \sim 1.5$
and a high energy cutoff at $\sim$100 keV .
This has been interpreted as inverse Comptonization of some soft seed photons by hot Maxwellian electrons (e.g., Shapiro et al. 1976; Sunyaev \& Tr\"{u}mper 1979) in a geomatrically thick and optically thin accretion flow, or ``corona", 
in which advection may be operating (e.g., Ichimaru 1977; Narayan \& Yi 1995).

A scenario of the LHS,  called  ``disk-corona" view (e.g., Liang et al. 1977), assumes that the system still harbors a standard disk, which provides the seed photons
for Comptonization.
Although the ``disk-corona" model became a standard interpretation of the LHS of BHBs, the overall geometry of the
disk and the corona remained poorly understood, including in particular, whether the geometrically thin disk extends to the last stable orbit 
(e.g., Miller et al. 2006a,b; Rykoff et al. 2007) or is truncated at some larger radii (e.g., Ebisawa et al. 1996; Poutanen et al. 1997; Gierlinski et al. 2008; Nowak et al. 2011).
Equally unsettled is the interpretation of the rapid variation (e.g., Miyamoto et al. 1989) in terms of the disk-corona view.
It was not clear, either, how these components evolve as the system approaches the HSS through the LHS.
These limitations have been caused mainly by observational difficulties, to obtain high-quality spectra on short time scales, and over an energy range wide enough 
to determine simultaneously and precisely both the disk emission around $\sim$1 keV and hard continuum shapes including the high energy cutoff energy of $\sim$ 100 keV.

In 2005 October, a broad-band Suzaku observation of Cyg X-1 in the LHS was conducted for an exposure of 17 ks (Makishima et al. 2008; hereafter Paper I).
The obtained 0.7--300 keV spectrum has been reproduced successfully by assuming that the disk is truncated at  $\gtrsim 15 R_{\rm G}$, and it provides the corona with the seed photons.
Here, $R_{\rm G}=G M_{\rm BH}/c^2$ is the gravitational radius, with $G$ the gravitational constant,
$M_{\rm{BH}}$ the black hole mass, and $c$ the speed of light.
The data also required a reflection hump with a solid angle of $\Omega/2\pi \sim 0.4$ and a mildly broadened iron line with a breadth of $\sigma \sim 1$ keV;
this is consistent with the disk truncation at $\sim$ 15 $R\rm{_G}$.
The corona was inferred to be highly inhomogenious, expressed by a two-zone approximation invoking optical depths of $\sim 0.4$ and $\sim 1.5$,
and the same electron temperature of $\sim$ 100 keV. In addition,
variations on a time scale of 1 s were studied in terms of Comptonized emission using intensity-sorted spectral analysis.
It was then indicated that the X-ray intensity flares up, on time scales of $\sim$ 1 s, when a larger fraction of the disk is covered by the corona, while the disk itself stays almost constant.
In other words, the rapid X-ray flickering of Cyg X-1 in the LHS is primarily generated by fluctuations of the number of disk photons intercepted by the corona.
This view is consistent with the one, based on the almost energy-independent fractional rms variation spectrum, that the variation in the LHS is generated by fluctuations in the seed photon input
(Gierli\'{n}ski \& Zdziarski 2005; Gierli\'{n}ski et al. 2010). Actually, figure 8(b) of Paper I  (high-phase to low-phase spectral ratio), which is almost equivalent to such rms spectra, is rather flat, except in
$\lesssim$ 1 keV where the ratio decreases due to the dominance of the disk emission (Yamada 2011).
Thus, the results of Paper I reinforced the disk-corona picture, and quantified its details.

Given this snap-shot result, our next task is to clarify long-term evolution of the disk-corona configuration,
which is a key to understanding the transition from the LHS to the HSS.
For this purpose, we analyzed 25 Suzaku data sets of Cyg X-1 acquired over 2005--2009. 
Among the observations, the spectral and timing properties in hard X-rays (10--200 keV) showed significant changes,
even though the source remained in the LHS throughout.
In the present paper, we limit our analysis to the data obtained with the Hard X-ray Detector (HXD; Takahashi et al. 2007; Kokubun et al. 2007).
Results with the X-ray Imaging Spectrometer (XIS; Koyama et al. 2007) will be reported elsewhere, after the data are fully corrected for pile up effects.

\section{Observation and Data Reduction}

\subsection{Suzaku observations}
\label{sec:suzakuobs}
So far, Cyg X-1 has been observed with Suzaku on 25 occasions from 2005 October 5 to 2009 December 17,
with a total exposure of 446 ks.
The exposure of individual observations ranges from 3 ks to 45 ks, with an average of 18 ks.
Intervals between a pair of consecutive observations range from $\sim$2 days to $\sim$1 year.
A log of these observations is given in table \ref{tab:obslog}.
The optical axes of the XIS and the HXD differ slightly by 3.5$'$, and
the first observation was performed at the XIS nominal position,
while the others at the HXD one.

The data to be analyzed in the present paper were obtained from the HXD,
which consists of Si PIN photo-diodes (HXD-PIN) covering 10--70 keV and GSO scintillation counters (HXD-GSO) covering 50--600 keV.
Sometimes the exposure is shorter than was aimed, due to increased electronics noise in HXD-PIN which reduced the live time.
The exposures after correcting for dead times are shown in table \ref{tab:obslog}. 
Among them, Observation 1 is the one which was conducted in the Performance Verification phase
by the Suzaku team, and already reported in Paper I.
The utilized HEADAS version is 6.9, unless otherwise stated.

\begin{figure}
\begin{center}
\FigureFile(80mm,80mm){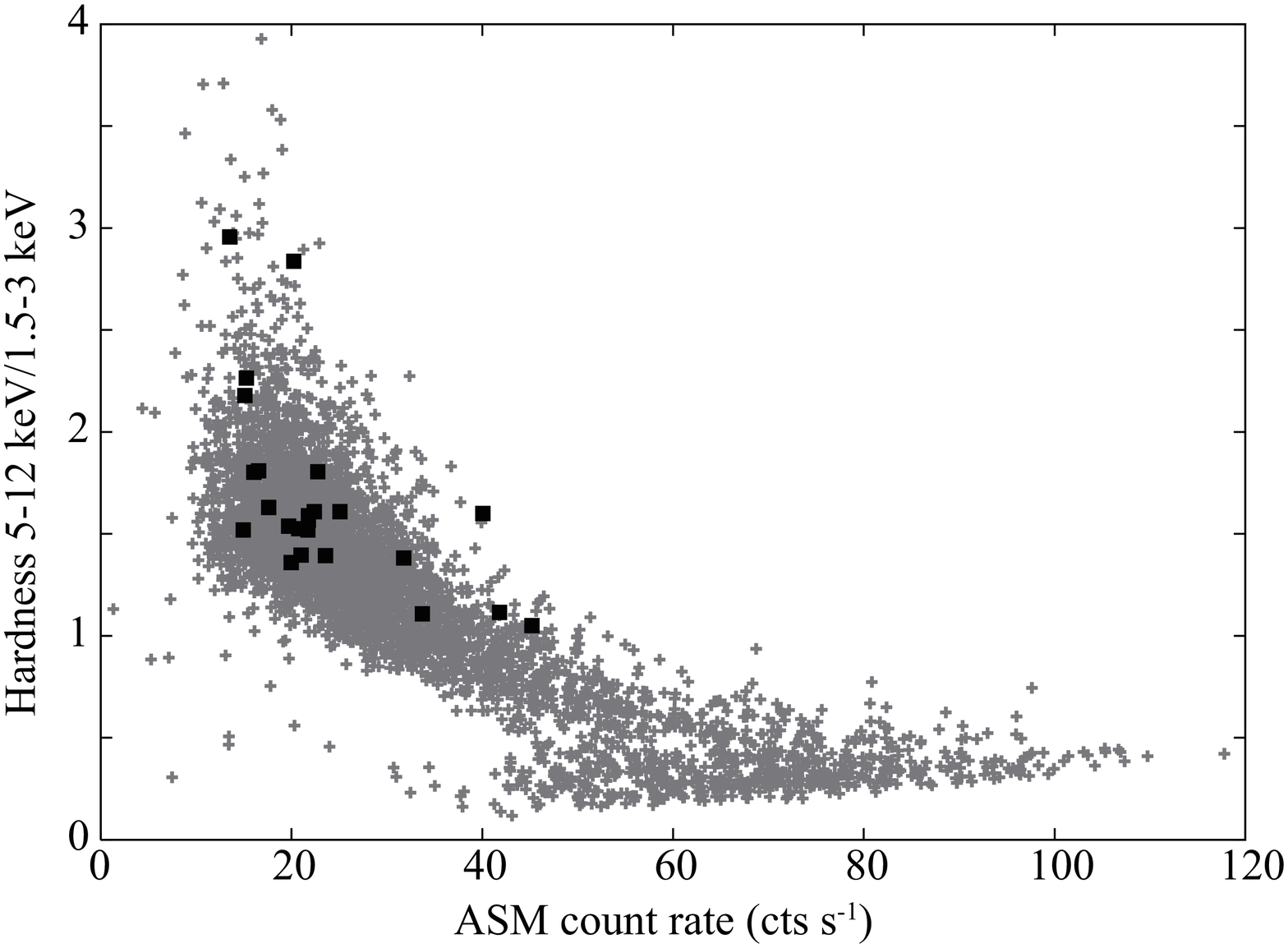}
\end{center}
\caption{
A scatter plot between the 1.5--12.0 keV count rate and the 5.0--12.0 keV vs 1.5--3.0 keV hardness ratio of Cyg X-1, obtained with the RXTE ASM over a period of 1996 December to 2010 September.
Each gray dot is an integration over one day. Black points indicate epochs of the 25 Suzaku observations.}
\label{fig:ASMCountAndHard}
\end{figure}

Figure \ref{fig:ASMCountAndHard} is a hardness-intensity diagram of Cyg X-1, constructed from long-term 5.0--12.0 keV and 1.5--3.0 keV data of the All Sky Monitor (ASM) onboard Rossi X-ray Timing Explorer.
As represented by black dots in figure \ref{fig:ASMCountAndHard}, the present Suzaku observations were all conducted when Cyg X-1 was in the typical LHS.
Hereafter, we designate the 1.5-12.0 keV ASM intensity as $C_{\rm{ASM}}$ and use it as a measure of the soft X-ray intensity of Cyg X-1,
because the XIS data still need extensive corrections for the pile up effects. This $C_{\rm{ASM}}$ is also considered as a rough measure
of the mass accretion rate.

\subsection{Data Reduction and Background Subtraction}

We processed the 25 data sets in the same manner. Specifically,
the PIN and GSO events were selected by criteria of elevation angle $\geq$ $5^{\circ}$, cutoff rigidity $\geq$ 6 GV, and 500 s after
and 180 s before the South Atlantic Anomaly.
The unscreened GSO data were reprocessed using $\tt{hxdpi}$ and $\tt{hxdgrade}$ to take into account the recent updates in the GSO calibration
(Yamada et al. 2011, in press), while the PIN data did not need reprocessing.

Since the HXD has neither imaging capability nor an offset detector, the background contribution to the on-source data
must be estimated by simulations. The HXD background is dominated by the non-X-ray background, hereafter NXB.
That of PIN was simulated by ``PINUDLCUNIT" model (Fukazawa et al. 2009). Using the PIN upper-discriminator hit rate as a major NXB indicator,
this model can reproduce the 10--70 keV NXB to an accuracy of 3\% or better for an exposure longer than 10 ks. The GSO NXB was simulated
by ``LCFITDT" model (Fukazawa et al. 2009), which tries to reproduce the GSO background  in 32 energy bands by analyzing long-term ($\sim$1 month) light curves therein.
This method achieves an accuracy of $\sim$ 1\% in background reproduction when the exposure is longer than 10 ks. 
The cosmic X-ray background (Boldt et al. 1987) can be ignored in our analysis, since Cyg X-1 is so bright an object.

\subsection{Light Curves of the HXD Data}

\begin{figure*}
\begin{center}
\FigureFile(170mm,170mm){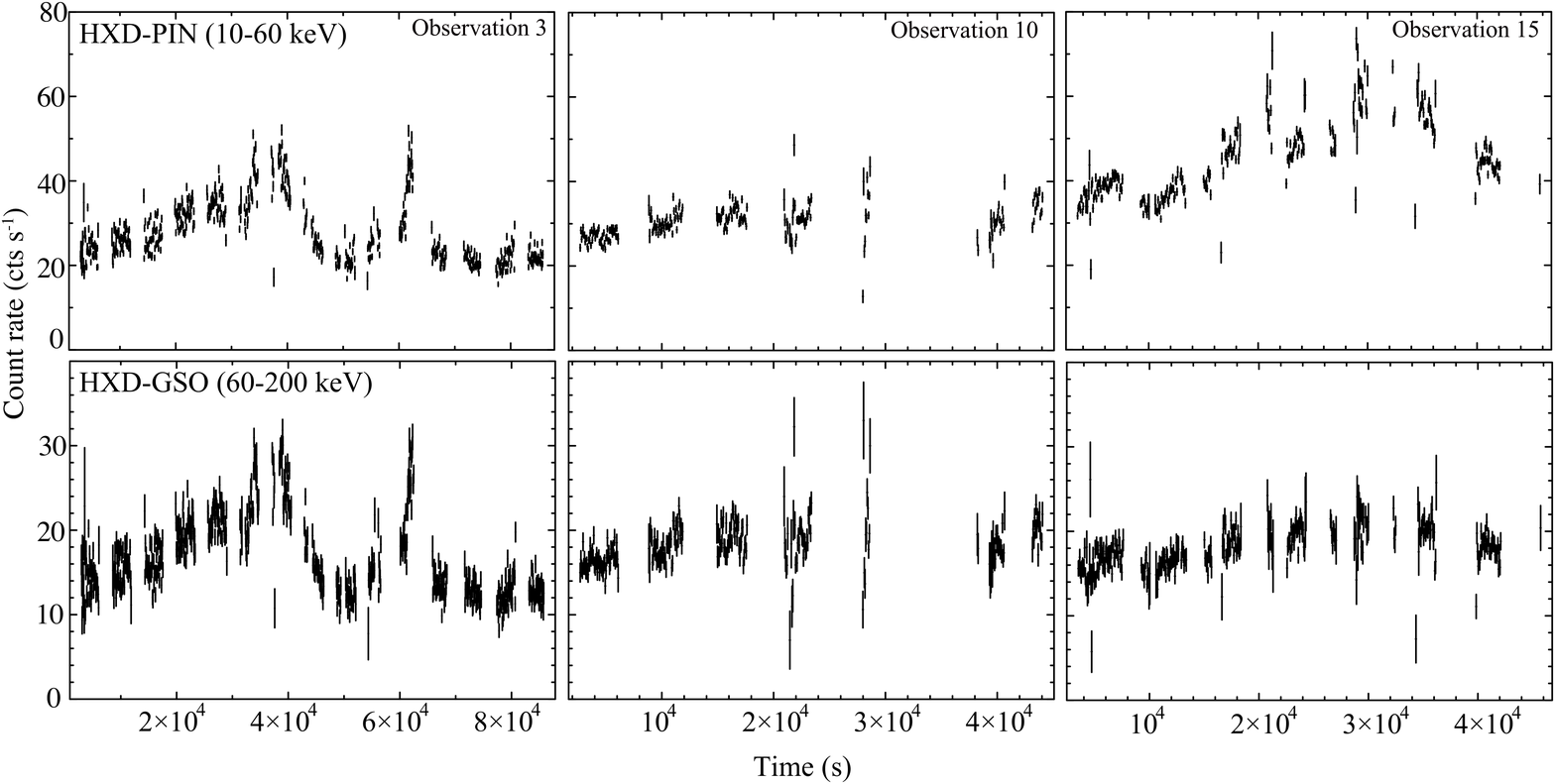}
\end{center}
\caption{ Background-subtracted 64-s bin light curves of Cyg X-1 with HXD-PIN (10--60 keV;top panels) and HXD-GSO (60--200 keV;bottom panels) from three representative observations. 
Left two panels are for Observation 3, middle two panels for Observation 10, while the right two panels Observation 15.}
\label{fig:lightcurve}
\end{figure*}

We have selected three typical observations, 3,10, and 15 in table \ref{tab:obslog}, and show their two-band HXD light curves in figure \ref{fig:lightcurve} in the increasing order of the PIN count rate (which happens to be also the increasing order of the observation number).
After subtracting the background and applying dead time corrections, the HXD-PIN (10--60 keV) and the HXD-GSO (60--200 keV) count rates in these observations
were typically 20--50 cts s $^{-1}$ and 20--30 cts s $^{-1}$, respectively. The light curves reveal intensity variations by 20--30\% on time scales of $\gtrsim$ 200 s, although those on shorter time scales are
difficult to resolve in these plots.
Hereafter, we use these three observations as the representatives of the data acquired at different source intensities.

\begin{figure*}
\begin{center}
\FigureFile(170mm,170mm){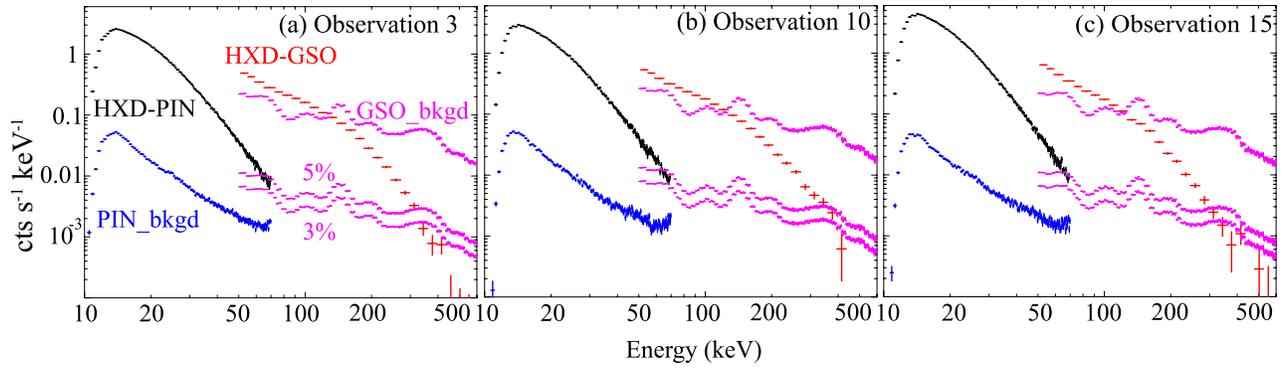}
\end{center}
\caption{
Time-averaged and response-inclusive spectra of HXD-PIN (black) and HXD-GSO (red) from the three representative observations, show after subtracting the simulated background data denoted as PIN\_bkgd (blue)
and GSO\_bkgd (magenta), respectively. Also shown in magenta are 3\% and 5\% levels of the GSO background. Panels (a), (b) and (c) correspond to the three epochs selected in figure \ref{fig:lightcurve}.}
\label{fig:specwithbgd}
\end{figure*}

Background-subtracted and dead-time-corrected spectra of these three observations are shown in figure \ref{fig:specwithbgd}, in the response folded form. In the PIN band, the background subtracted
signals in these three observations exceed the NXB by a factor of $\sim$ 5, even at 60 keV. In the GSO band, the source signal is detected up to 300--400 keV, since it therein exceeds 3\% of the background while
the typical background uncertainty is $\sim$ 1\% (Fukazawa et al. 2009).



\begin{table*}
\caption{An observation log of Cyg X-1.}
\begin{center}
\small
\begin{minipage}{14cm}
\begin{tabular*}{14cm}
{ccccccccc}
\hline
\hline
No. &Date & MJD & ObsID & Exposure & D.V.\footnotemark[*] & Pointing & epoch\footnotemark[$\dagger$] &$C_{\rm{ASM}}$\\
&&&&(ks)&&&& (cts s$^{-1}$)\\
\hline
1&2005/10/05 & 53648 & 100036010 & 18.1 & 2.1 & XIS & 1 & 28.6 \\
2&2006/10/30 & 54038 & 401059010 & 27.7 & 2.0 & HXD & 3 & 21.0 \\
3&2007/04/30 & 54220 & 402072010 & 40.2 & 2.1 & HXD & 3 & 14.9 \\
4&2007/05/17 & 54237 & 402072020 & 32.6 & 2.0 & HXD & 3 & 13.5 \\
5&2008/04/18 & 54574 & 403065010 & 29.0 & 2.2 & HXD & 4 & 15.3 \\
6&2009/04/03 & 54924 & 404075010 & 15.5 & 2.3 & HXD & 5 & 21.7 \\
7&2009/04/08 & 54929 & 404075020 & 13.1 & 2.3 & HXD & 5 & 21.8 \\
8&2009/04/14 & 54935 & 404075030 & 13.0 & 2.3 & HXD & 5 & 17.6 \\
9&2009/04/23 & 54944 & 404075040 & 16.3 & 2.3 & HXD & 5 & 19.7 \\
10&2009/04/28 & 54949 & 404075050 & 12.4 & 2.3 & HXD & 5 & 20.8 \\
11&2009/05/06 & 54957 & 404075060 & 15.3 & 2.3 & HXD & 5 & 22.4 \\
12&2009/05/20 & 54971 & 404075080 & 17.5 & 2.4 & HXD & 5 & 23.6 \\
13&2009/05/25 & 54976 & 404075090 & 16.1 & 2.4 & HXD & 5 & 31.8 \\
14&2009/05/29 & 54980 & 404075100 & 25.7 & 2.4 & HXD & 5 & 33.7 \\
15&2009/06/02 & 54984 & 404075110 & 15.0 & 2.4 & HXD & 5 & 45.2 \\
16&2009/06/04 & 54986 & 404075120 & 6.8 & 2.4 & HXD & 5 & 41.8 \\
17&2009/10/21 & 55125 & 404075130 & 15.2 & 2.4 & HXD & 6 & 20.0 \\
18&2009/10/26 & 55130 & 404075140 & 20.0 & 2.4 & HXD & 6 & 16.1 \\
19&2009/11/03 & 55138 & 404075150 & 14.1 & 2.4 & HXD & 6 & 16.6 \\
20&2009/11/10 & 55145 & 404075160 & 18.3 & 2.4 & HXD & 6 & 15.1 \\
21&2009/11/17 & 55152 & 404075170 & 18.7 & 2.4 & HXD & 6 & 20.2 \\
22&2009/11/24 & 55159 & 404075180 & 11.4 & 2.4 & HXD & 6 & 21.8 \\
23&2009/12/01 & 55166 & 404075190 & 13.2 & 2.4 & HXD & 6 & 25.1 \\
24&2009/12/08 & 55173 & 404075200 & 17.4 & 2.4 & HXD & 6 & 22.8 \\
25&2009/12/17 & 55182 & 404075070 & 0.3 & 2.4 & HXD & 6 & 40.1 \\
\hline
\end{tabular*}
\footnotetext[*]{D.V. stands for data version of the HXD data which differ in calibration and data process.}
\footnotetext[$\dagger$]{Epoch is a period in which the same PIN response is applied.}
\end{minipage}
\label{tab:obslog}
\end{center}
\end{table*}

\section{Spectral Analysis}

\subsection{The HXD Hardness Ratio}

As a first-cut characterization of the HXD spectra taken in the 25 observations, we calculated ratios of the 60--200 keV
average count rate detected by HXD-GSO in each observation over that in the 10--60 keV by HXD-PIN, with the NXB subtracted from both.
This quantity, called HXD hardness ratio, can be used as an indicator of spectral hardness in the hard X-ray band.

\begin{figure}[h]
\begin{center}
\FigureFile(85mm,120mm){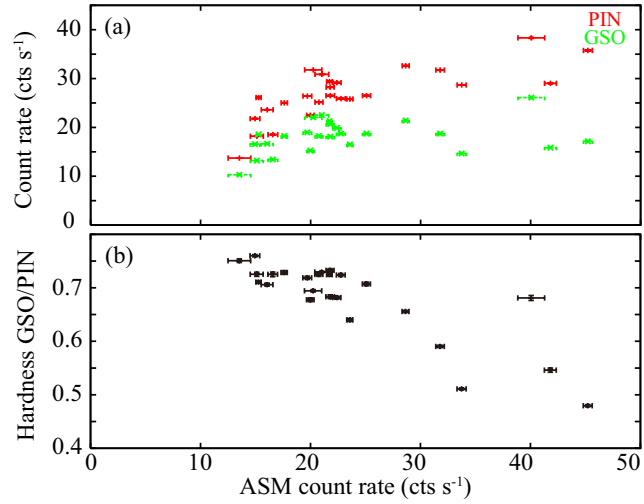}
\end{center}
\caption{(a) The count rate of PIN and GSO, and (b) the HXD hardness ratio (see text) of Cyg X-1 plotted against $C_{\rm{ASM}}$.}
\label{fig:HXDhardness}
\end{figure}

As figure \ref{fig:HXDhardness} shows, the obtained hardness ratio is clearly anti-correlated with $C_{\rm{ASM}}$; this is mainly due to negative $C_{\rm{ASM}}$ dependence of the PIN count rate, while the
GSO count rate is rather constant.
This result extends the intensity versus hardness anti-correlation seen in the ASM range (figure \ref{fig:ASMCountAndHard}) to the HXD band.

\subsection{Spectral Fitting}

For detailed spectral analysis, we prepared response files for HXD-PIN and HXD-GSO using xspec12. As listed in table 1, the HXD-PIN responses were selected according
to ``epochs" of the observation, which are updated when lower-discriminator levels and/or high-voltage values of HXD-PIN are altered (Takahashi et al. 2007; Kokubun et al. 2007).
The source positions in the field of view, explained in subsection 2.1, also affect the responses. We applied $\tt{ae\_hxd\_pinXXnomeY\_20100107.rsp}$ to the PIN spectra and $\tt{ae\_hxd\_gsoXXnom\_20110107.rsp}$ 
and $\tt{ae\_hxd\_gsoXXnom\_crab\_20100526.arf}$ to those of HXD-GSO. Here, $\tt{XX}$ and $\tt{Y}$ specify the source positions (XIS nominal as ``$\tt{xi}$" and HXD nominal as ``$\tt{hx}$")
and the epoch (``$\tt{1}$" to ``$\tt{6}$"), respectively. The arf file for HXD-GSO is a correction file to adjust the normalization between HXD-PIN and HXD-GSO, and
to trim high energy responses (Takahashi et al. 2007).
The PIN data below 14 keV were discarded since electrical noises can contaminate the data up to this energy. The GSO data below 70 keV and above 400 keV
were also excluded from fitting, because of large response uncertainties and poor statistics, respectively. 
Because Cyg X-1 was bright, we added a systematic errors of 1\%, representing the NXB reproducibility (Fukazawa et al. 2009), to all bins.

\begin{figure}
\begin{center}
\FigureFile(80mm,80mm){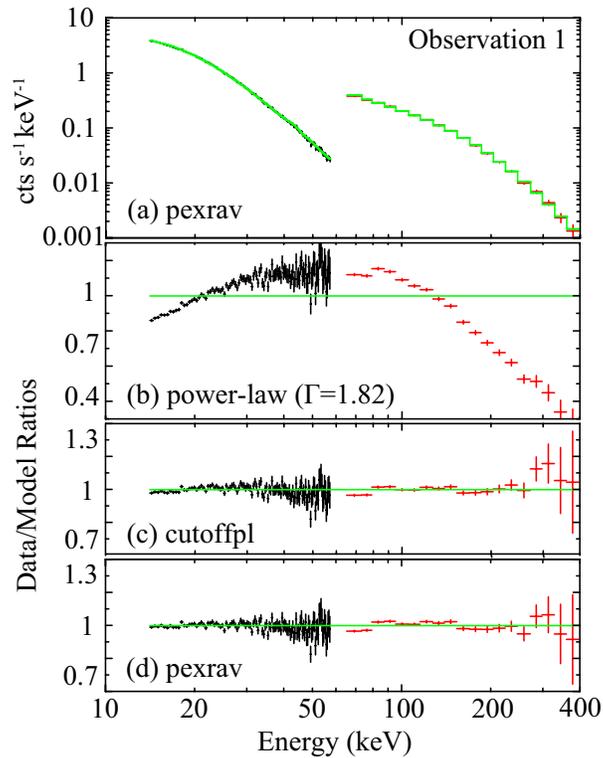}
\end{center}
\caption{(a) The response-folded PIN and GSO spectra from the first observation, compared with the best-fit pexrav model. Panels (b), (c) and (d) shows the data divided the best-fit $\tt{powerlaw}$ model, $\tt{cutoffpl}$ model, and the $\tt{pexrav}$ model (corresponding to panel a), respectively. }
\label{fig:empirical}
\end{figure}

In order to grasp a spectral shape and reconfirm the analysis procedure in Paper I under a condition of using only the HXD data, we fitted the time-averaged HXD spectrum of the first observation with a single power-law model. 
The interstellar absorption was expressed by $\tt{wabs}$ in xspec, and its column density was fixed at $N_{\rm{H}} = 6 \times 10^{21}~{\rm{cm^{-2}}}$ after Paper I
since its effects are minor in the HXD energy band. The best fit photon index was obtained as $\Gamma =$ 1.78. However, the fit was far from being acceptable,
with $\chi^2\slash{\nu}$= 4262/132, because the data are as shown in figure \ref{fig:empirical} (b) more convex than the model.
This agrees with a number of past reports,
including Paper I.
Therefore, we substituted a cutoff power-law model for the single power-law model. The fit was improved to $\chi^2\slash{\nu}$=170/131, with $\Gamma =$1.39
and the exponential cutoff energy $E_{\rm{cut}}$=162 keV. However, the fit is not yet fully acceptable, because of some excess in the 20-40 keV range as seen in figure \ref{fig:empirical} (c).

To account for this structure, we introduced a reflection component after Paper I. The utilized model is again $\tt{pexrav}$ (Magdziarz \& Zdziarski 1995), which takes into account reflection from 
cold matter and a spectral cutoff. Like in Paper I, we fixed the inclination angle of the reflection plane at 45$^\circ$. As shown in figure \ref{fig:empirical} (d), the fit became fully acceptable with $\Gamma =$1.44,
 $E_{\rm{cut}}$=158 keV, a reflection solid angle $\Omega \slash{2\pi}$=0.14 and $\chi^2\slash{\nu}$=148/130.

Now that an acceptable fit was achieved in the HXD range by an empirical cutoff power-law model including reflection, the next step is to employ the more meaningful
description of the spectrum, namely unsaturated inverse Comptonization processes (Paper I). For this purpose, we utilized $\tt{compPS}$ model
(Poutanen \& Svensson 1996), which takes into account not only relativistic Comptonization correctly but also reflection by a cold matter.
Free parameters of the model in our analysis are electron temperature $kT_{\rm{e}}$, optical depth $\tau$, reflection solid angle $\Omega \slash{2\pi}$ and normalization.
The seed photon temperature was assumed to be 0.2 keV (Paper I), since this parameter cannot be determined by the present data
which are limited to above 10 keV. We assumed a spherical geometry in the spectral analysis.
Then, as listed in table \ref{tab:fitpara}, the $\tt{CompPS}$ fit to Observation 1 has become fully acceptable, but has yielded the parameters
which are not fully consistent with those of the higher $\tau$ component derived in Paper I. The reason of these deviations is discussed in section 5.

\begin{figure*}
\begin{center}
\FigureFile(170mm,170mm){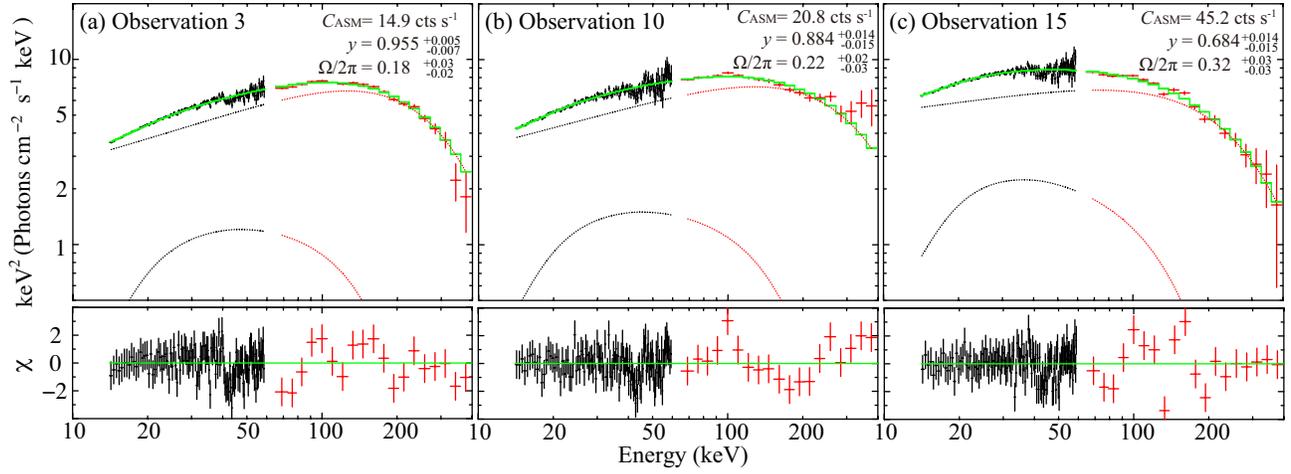}
\end{center}
\caption{The same HXD spectra of Cyg X-1 as in figure \ref{fig:specwithbgd}, presented in the deconvolved ${\nu}{F}{\nu}$ form. The best-fit model and the fit residuals are also shown, where the Compton component and the reflection component are displayed separately.}
\label{fig:CygSpec}
\end{figure*}

\begin{figure}
\begin{center}
\FigureFile(73mm,80mm){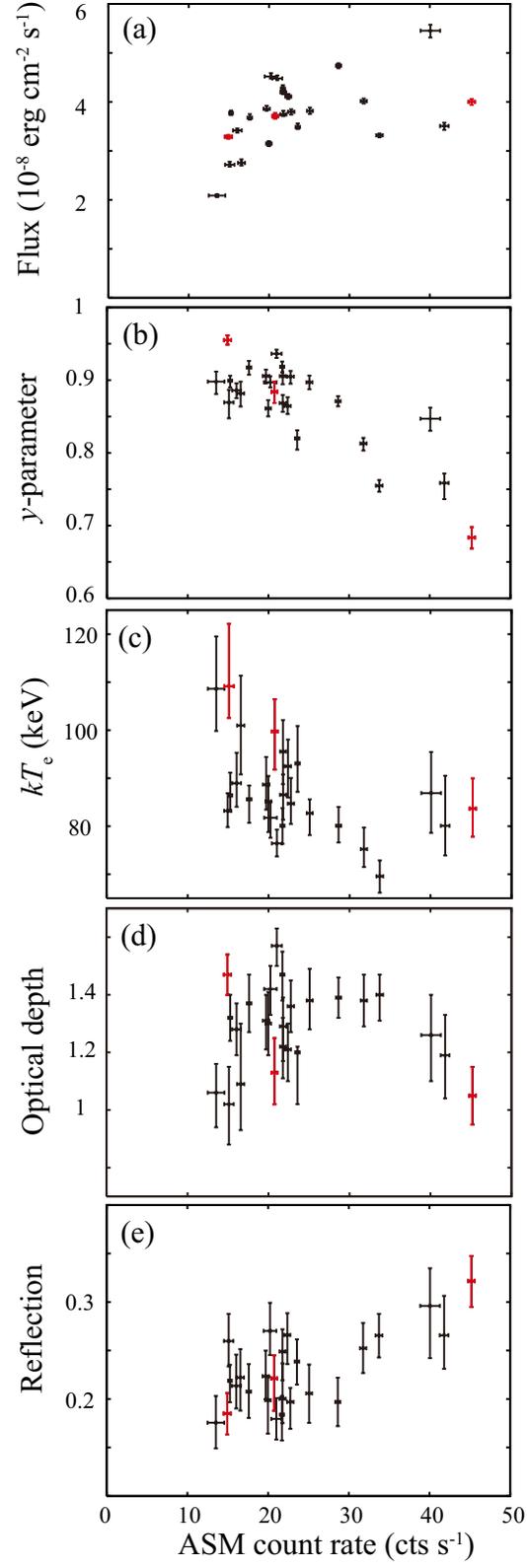}
\end{center}
\caption{The Compton ($\tt{CompPS}$) parameters from the 25 observations of Cyg X-1, plotted against $C_{\rm{ASM}}$. (a) The 10--400 keV HXD model flux, (b) $\it{y}$, (c) $T_{\rm{e}}$, (d) $\tau$, and (e) $\Omega \slash{2\pi}$. The three representatives are shown in red.}
\label{fig:specpara}
\end{figure}

We next applied this model to the same three observations as in figure \ref{fig:lightcurve}, and obtained the results
shown in figure \ref{fig:CygSpec} and table \ref{tab:fitpara}. Since the three fits were all successful,
we further applied the same model to the remaining observations, and again obtained acceptable fits in all of them (table \ref{tab:fitpara}).
In figure \ref{fig:specpara}, the obtained parameters are expressed as a function of  $C_{\rm{ASM}}$. From figure \ref{fig:specpara} (a), the HXD flux 
is seen to increase with $C_{\rm{ASM}}$, at least up to $C_{\rm{ASM}}$ $\sim$ 20 cts s$^{-1}$.  The values of $kT_{\rm{e}}$ = 70--100 keV and $\tau$ = 1--1.5 are consistent with those reported in previous works (e.g. Shapiro et al. 1976; Sunyaev \& Tr\"{u}mper 1979; Sunyaev \& Titarchuk 1980; Gierli\'{n}ski et al. 1997; Di Salvo et al. 2001; Frontera et al. 2001a; Zdziarski \& Gierli\'{n}ski 2004; Ibragimov et al. 2005). 

In figure \ref{fig:specpara}, neither $kT_{\rm{e}}$ nor $\tau$ exhibits significant dependence on $C_{\rm{ASM}}$. However, as expected from the spectral slope change observed in figure \ref{fig:CygSpec}, a clear anti-correlation is revealed in figure \ref{fig:specpara} (b) between
$C_{\rm{ASM}}$ and the $y$-parameter, $y \equiv 4\tau kT_{\rm{e}}/(m_{\rm{e}}c^2)$, where $k$ is the Boltzmann constant, $m_{\rm{e}}$ is the electron mass, and $c$ is the light velocity.
Moderate values of reflection, $\Omega \slash{2\pi}$ = 0.20--0.35, are consistent with previous works, considering that Suzaku covered near the hardest end of the LHS
(figure \ref{fig:ASMCountAndHard}). As already suggested by figure \ref{fig:CygSpec}, $\Omega \slash{2\pi} $ depends positively on $C_{\rm{ASM}}$.
Similar positive relations between $\Gamma$ and  $\Omega \slash{2\pi} $ have been observed not only from Cyg X-1 (Gilfanov et al. 1999; Ibragimov et al. 2005; Titarchuk et al. 2007), but also from other BHBs and Seyfert galaxies (Zdziarski et al. 1999).

\begin{table*}
\caption{Parameters obtained by fitting the HXD spectra with a single \tt{CompPS} model.}
\begin{center}
\small
\begin{tabular*}{15.5cm}
{cccccccc}
\hline
\hline
No. & 10--400 keV flux &$y$ & $ kT_{\rm{e}}$ & $\tau$ & $\Omega/{2\pi}$ & $\chi^{2}_{\nu}(\nu)$ \ \& \ d.o.f. & $C_{\rm{ASM}}$\\
&(10$^{-8}$ erg cm$^{-2}$ s$^{-1}$)&&(keV)&&&&(cts s$^{-1}$)\\
\hline
1	&	4.74$_{-0.05}^{+0.05}$	&	0.871$_{-0.007}^{+0.007}$	&	80.1$_{-3.5}^{+3.9}$	&	1.39$_{-0.07}^{+0.07}$	&	0.20$_{-0.03}^{+0.02}$	&	1.07(130)	&28.6$\pm$0.3	\\
2	&	4.48$_{-0.03}^{+0.06}$	&	0.936$_{-0.005}^{+0.006}$	&	76.4$_{-2.7}^{+2.9}$	&	1.57$_{-0.07}^{+0.06}$	&	0.18$_{-0.02}^{+0.02}$	&	1.00(134)	&21.0$\pm$0.6	\\
3	&	3.29$_{-0.04}^{+0.03}$	&	0.955$_{-0.006}^{+0.006}$	&	83.2$_{-3.4}^{+3.7}$	&	1.47$_{-0.07}^{+0.07}$	&	0.18$_{-0.02}^{+0.03}$	&	1.15(134)	&14.9$\pm$0.4	\\
4	&	2.09$_{-0.03}^{+0.03}$	&	0.898$_{-0.017}^{+0.014}$	&	108.6$_{-8.8}^{+10.9}$	&	1.06$_{-0.12}^{+0.10}$	&	0.18$_{-0.03}^{+0.02}$	&	0.74(134)	&13.5$\pm$1.0	\\
5	&	3.77$_{-0.04}^{+0.05}$	&	0.899$_{-0.009}^{+0.007}$	&	86.4$_{-3.4}^{+4.8}$	&	1.32$_{-0.08}^{+0.08}$	&	0.22$_{-0.02}^{+0.02}$	&	1.09(134)	&15.3$\pm$0.2	\\
6	&	4.27$_{-0.05}^{+0.07}$	&	0.918$_{-0.007}^{+0.007}$	&	80.1$_{-3.8}^{+3.7}$	&	1.47$_{-0.09}^{+0.08}$	&	0.18$_{-0.02}^{+0.02}$	&	1.08(134)	&21.7$\pm$0.3	\\
7	&	4.19$_{-0.03}^{+0.07}$	&	0.905$_{-0.011}^{+0.012}$	&	95.6$_{-6.8}^{+6.5}$	&	1.22$_{-0.11}^{+0.10}$	&	0.25$_{-0.03}^{+0.02}$	&	1.17(134)	&21.8$\pm$0.4	\\
8	&	3.68$_{-0.04}^{+0.07}$	&	0.917$_{-0.010}^{+0.009}$	&	85.6$_{-4.9}^{+2.9}$	&	1.37$_{-0.10}^{+0.10}$	&	0.21$_{-0.03}^{+0.03}$	&	1.20(134)	&17.6$\pm$0.3	\\
9	&	3.86$_{-0.04}^{+0.06}$	&	0.906$_{-0.010}^{+0.008}$	&	88.7$_{-5.3}^{+5.7}$	&	1.31$_{-0.10}^{+0.09}$	&	0.22$_{-0.02}^{+0.03}$	&	1.11(134)	&19.7$\pm$0.4	\\
10	&	3.71$_{-0.05}^{+0.06}$	&	0.884$_{-0.015}^{+0.014}$	&	99.7$_{-7.9}^{+6.7}$	&	1.13$_{-0.11}^{+0.12}$	&	0.22$_{-0.03}^{+0.02}$	&	1.07(134)	&20.8$\pm$0.4	\\
11	&	4.11$_{-0.05}^{+0.05}$	&	0.864$_{-0.011}^{+0.012}$	&	92.5$_{-6.5}^{+5.5}$	&	1.21$_{-0.11}^{+0.09}$	&	0.27$_{-0.03}^{+0.02}$	&	1.06(134)	&22.4$\pm$0.4	\\
12	&	3.49$_{-0.04}^{+0.07}$	&	0.820$_{-0.016}^{+0.011}$	&	93.1$_{-6.0}^{+7.8}$	&	1.20$_{-0.18}^{+0.02}$	&	0.24$_{-0.03}^{+0.02}$	&	1.00(134)	&23.6$\pm$0.3	\\
13	&	4.01$_{-0.04}^{+0.06}$	&	0.813$_{-0.010}^{+0.008}$	&	75.2$_{-3.7}^{+4.5}$	&	1.38$_{-0.09}^{+0.09}$	&	0.25$_{-0.02}^{+0.03}$	&	0.91(134)	&31.8$\pm$0.4	\\
14	&	3.32$_{-0.04}^{+0.03}$	&	0.755$_{-0.008}^{+0.008}$	&	69.6$_{-3.5}^{+3.3}$	&	1.40$_{-0.09}^{+0.07}$	&	0.27$_{-0.03}^{+0.02}$	&	1.38(134)	&33.7$\pm$0.4	\\
15	&	4.00$_{-0.06}^{+0.05}$	&	0.684$_{-0.015}^{+0.014}$	&	83.7$_{-5.9}^{+6.3}$	&	1.05$_{-0.10}^{+0.10}$	&	0.32$_{-0.03}^{+0.03}$	&	1.16(134)	&45.2$\pm$0.4	\\
16	&	3.51$_{-0.08}^{+0.06}$	&	0.759$_{-0.023}^{+0.013}$	&	80.1$_{-6.2}^{+10.4}$	&	1.19$_{-0.15}^{+0.14}$	&	0.27$_{-0.04}^{+0.04}$	&	1.04(125)	&41.8$\pm$0.5	\\
17	&	3.15$_{-0.05}^{+0.05}$	&	0.862$_{-0.012}^{+0.010}$	&	85.2$_{-6.4}^{+5.3}$	&	1.30$_{-0.11}^{+0.11}$	&	0.20$_{-0.04}^{+0.02}$	&	1.21(133)	&20.0$\pm$0.3	\\
18	&	3.42$_{-0.05}^{+0.04}$	&	0.886$_{-0.011}^{+0.010}$	&	89.0$_{-5.0}^{+6.3}$	&	1.28$_{-0.09}^{+0.09}$	&	0.21$_{-0.02}^{+0.04}$	&	1.12(134)	&16.1$\pm$0.5	\\
19	&	2.75$_{-0.05}^{+0.09}$	&	0.882$_{-0.018}^{+0.016}$	&	101.0$_{-10.2}^{+10.4}$	&	1.09$_{-0.16}^{+0.21}$	&	0.22$_{-0.03}^{+0.03}$	&	0.94(129)	&16.6$\pm$0.4	\\
20	&	2.72$_{-0.04}^{+0.06}$	&	0.869$_{-0.022}^{+0.018}$	&	109.2$_{-6.7}^{+12.9}$	&	1.02$_{-0.14}^{+0.13}$	&	0.26$_{-0.03}^{+0.03}$	&	1.26(133)	&15.1$\pm$0.6	\\
21	&	4.52$_{-0.04}^{+0.07}$	&	0.897$_{-0.007}^{+0.008}$	&	81.8$_{-4.2}^{+3.3}$	&	1.42$_{-0.09}^{+0.08}$	&	0.27$_{-0.02}^{+0.03}$	&	1.11(134)	&20.2$\pm$0.8	\\
22	&	3.74$_{-0.03}^{+0.08}$	&	0.868$_{-0.012}^{+0.012}$	&	86.5$_{-5.1}^{+4.3}$	&	1.29$_{-0.12}^{+0.10}$	&	0.20$_{-0.02}^{+0.04}$	&	0.97(132)	&21.8$\pm$0.4	\\
23	&	3.82$_{-0.07}^{+0.06}$	&	0.897$_{-0.009}^{+0.009}$	&	82.7$_{-4.5}^{+2.9}$	&	1.38$_{-0.10}^{+0.11}$	&	0.21$_{-0.03}^{+0.03}$	&	0.94(134)	&25.1$\pm$0.4	\\
24	&	3.80$_{-0.06}^{+0.05}$	&	0.905$_{-0.010}^{+0.008}$	&	84.7$_{-4.2}^{+5.3}$	&	1.36$_{-0.09}^{+0.09}$	&	0.20$_{-0.03}^{+0.01}$	&	1.18(134)	&22.8$\pm$0.4	\\
25	&	5.45$_{-0.14}^{+0.13}$	&	0.847$_{-0.017}^{+0.015}$	&	86.9$_{-8.3}^{+8.6}$	&	1.26$_{-0.16}^{+0.14}$	&	0.30$_{-0.06}^{+0.03}$	&	1.00(118)	&40.1$\pm$1.2	\\

\hline
\end{tabular*}
\label{tab:fitpara}
\end{center}
\end{table*}

\section{Timing Analysis}

\subsection{Power Spectral Density}

By using $\tt{powspec}$ in XRONOS, we computed power spectral densities (PSDs) of all observations, separately on the 10--60 keV PIN events and on the 60--200 keV GSO events.
The NXB events, which account for $\sim50\%$ of the GSO data and $\sim2\%$ of the PIN data, were subtracted.
PSDs of Cyg X-1 in the LHS are known to have a nearly constant power density in frequencies below $\nu_{\rm{b}}$ $\sim$ 0.1 Hz, and becomes ``red" above $\nu_{\rm{b}}$
 (Miyamoto et al. 1989; Negoro et al. 2001; Titarchuk et al. 2006).
This $\nu_{\rm{b}}$ is referred to as a ``break frequency" and thought to have some relations to variation time scales of BHBs, perhaps a viscous time scale (Done et al, 2007; Ingram \& Done, 2011).
Therefore, our PSDs cover from $10^{-3}$ to 10 Hz, which fully contains the expected $\nu_{\rm{b}}$. Considering these, we employed a binning time of 0.05 s and a calculation interval of 8192 ($2^{13}$) bins. Considering our exposure efficiency of $\sim$ 50\%, any time interval which has less than 4096 bins was discarded. The PSDs are normalized so as to directly reflect the fractional rms variation, and are displayed after subtracting the expected white noise arising
from the counting statistics.

\begin{figure}
\begin{center}
\FigureFile(80mm,80mm){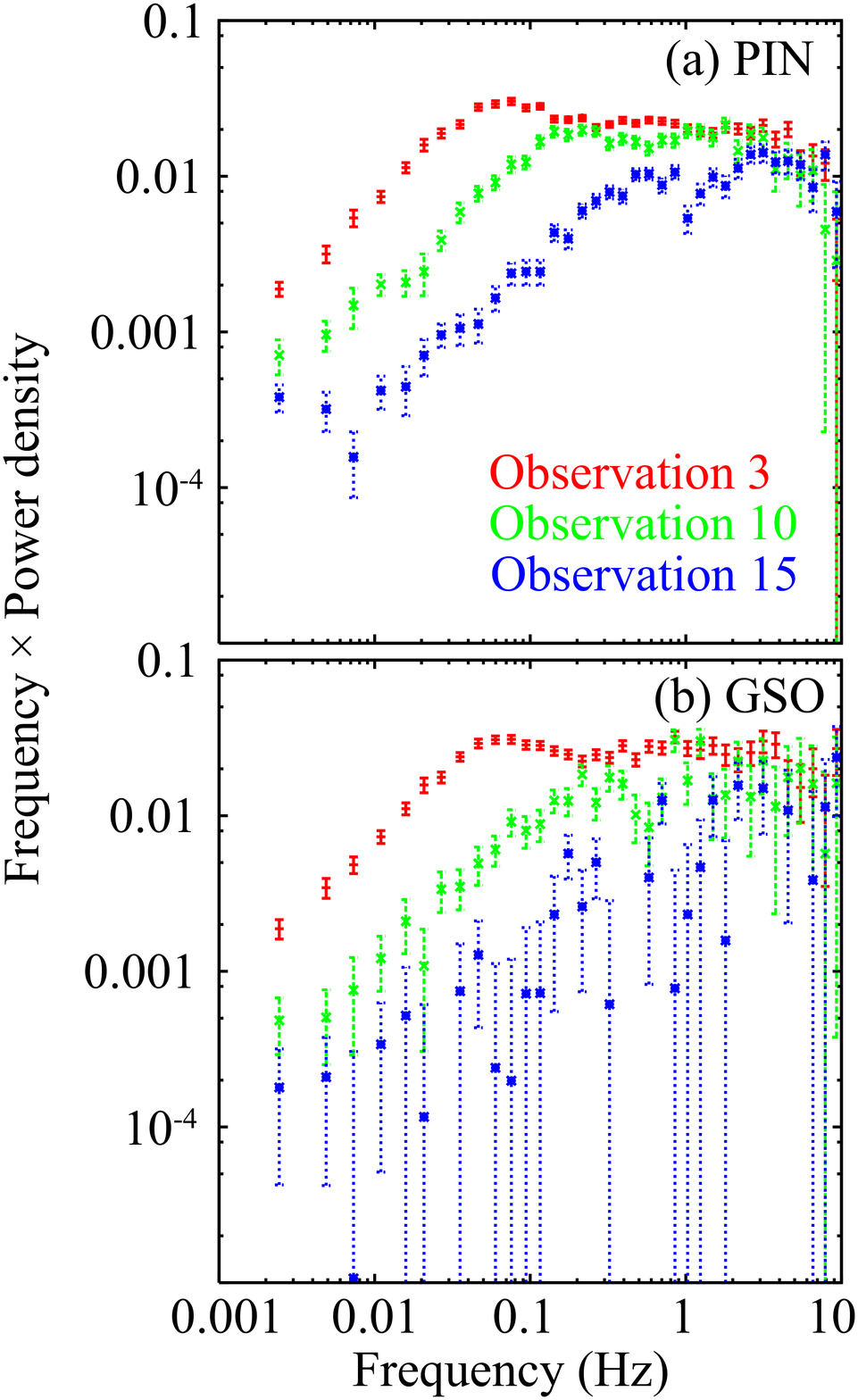}
\end{center}
\caption{Power spectral densities (PSDs), multiplied by the frequency $\nu$, of the PIN data (panel a) and the GSO data (panel b) from the three representative observations. Red shows Observation 3, green Observation 10 and blue Observation 15.}
\label{fig:PSD}
\end{figure}

Figure \ref{fig:PSD} shows PSDs multiplied by the frequency of the three representative observations which were selected in subsection 2.3. 
All of them exhibit shapes typical of this BHB, while the value of $\nu_{\rm{b}}$ and the power density below $\nu_{\rm{b}}$ are both seen to depend on the observation.
Interestingly, the PSD behavior above $\nu_{\rm{b}}$ is relatively the same, both in the slope and intensity, among the three data sets.

To quantify these PSDs, let us fit them with an empirical formula as

\begin{eqnarray}
\nu P(\nu)=\frac{a\nu}{(\nu_{\rm{b}}^2+\nu^2)^c}
\label{pnu}
\end{eqnarray}
where $\nu$ is the frequency, while the free parameters, $a$ and $c$, are normalization and the power-law index after the break, respectively.
According to RXTE PCA results, a PSD of Cyg X-1 in the LHS has roughly two break frequencies (Nowak et al. 1999; Pottschmidt et al. 2003).
Our $\nu_{\rm{b}}$ in equation \ref{pnu}, at 0.03--1 Hz in figure \ref{fig:PSD}, corresponds to their lower break frequency, while the effects of the higher break frequency are approximately accounted for by the parameter $c$. Only the PIN PSDs were fitted, because the GSO data give much poorly determined PSDs.

\begin{figure}
\begin{center}
\FigureFile(80mm,80mm){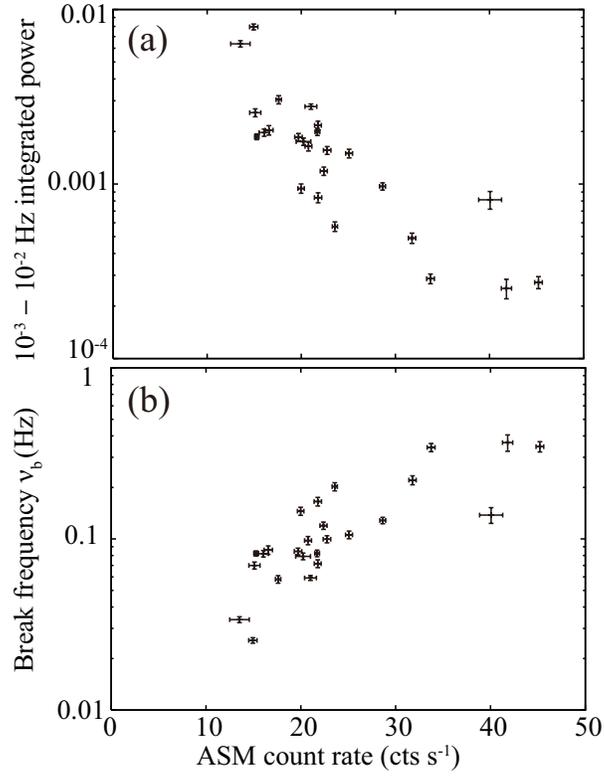}
\end{center}
\caption{Correlation between $C_{\rm{ASM}}$ and (a) power integrated over 10$^{-3}$--10$^{-2}$ Hz, and (b) the break frequency.}
\label{fig:PSDresult}
\end{figure}

We obtained acceptable fits with equation \ref{pnu} to all the 25 HXD-PIN PSDs, when we assume systematic errors of at least 3\%.
Figure \ref{fig:PSDresult} summarizes, as a function of $C_{\rm{ASM}}$, the obtained PSD parameters, where the low frequency power in panel (b) was calculated by integrating
$P(\nu)$ from 10$^{-3}$ Hz to 10$^{-2}$ Hz.
As $C_{\rm{ASM}}$ tripled, the value of ${\nu}_b$ increased from 0.03 to 0.3 Hz, and the variation power over 10$^{-3}$--10$^{-2}$ Hz decreased by a factor of $\sim$ 30 (or the
fractional variation decreased by a factor of $\sim$ 5). 
These mean that the time scale of variability shortens, and the variation amplitude in low frequencies decreases as $C_{\rm{ASM}}$ increases (or the energy spectrum softens).

\begin{figure}
\begin{center}
\FigureFile(80mm,80mm){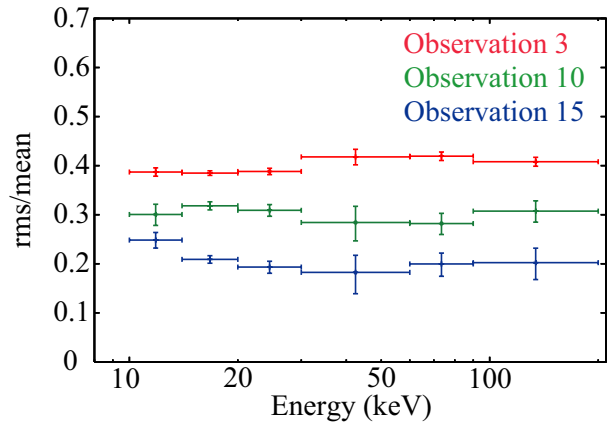}
\end{center}
\caption{The fractional rms variation spectra of the three representative observations.}
\label{fig:rmsspec}
\end{figure}

Since figure \ref{fig:PSDresult}(a) was integrated only over the very low 
frequency range (to avoid the changes in $\nu_{\rm b}$),
the implied variation amplitude is relatively low,  
i.e., 2--8\% of the mean.
In order to more properly estimate 
the variation amplitude and its energy dependence,
we divided the three representative HXD data into 6 energy bands, 
and calculated fractional rms variations 
by integrating the corresponding PSDs 
over the entire frequency range of 10$^{-3}$--10 Hz.
The Poisson noise contribution was subtracted. 
As figure \ref{fig:rmsspec} shows, 
the obtained  fractional rms variations are  20--40\%,
with a negative dependence on $C_{\rm ASM}$,
but little (or slightly negative) dependence on the energy.
These properties are consistent with previous 
measurements (e.g., Gierli{\'n}ski et al. 2010)


\subsection{Auto Correlation Function}

\label{sec:ACF}
We computed auto correlation functions (ACFs) of the 10--60 keV PIN data and the 60--200 keV GSO data,
as our second step for characterization of the time variability. Although an ACF and the corresponding PSD, being Fourier conjugate to each other, carry essentially the identical information,  the former works in time domain whereas the latter in frequency domain. Therefore, we may benefit by studying both of them, and examining them for their mutual consistency. The utilized tool is $\tt{autocor}$ in XRONOS, with a binning time 0.1 s and 4096 bins in each computation interval. In other words, we computed
the ACFs every 409.6 s with a time resolution of 0.1 s, and took their average. We chose to normalize the ACF values by the variances and the number of bins per interval,
and subtracted the Poisson noise contribution.

\begin{figure}
\begin{center}
\FigureFile(80mm,80mm){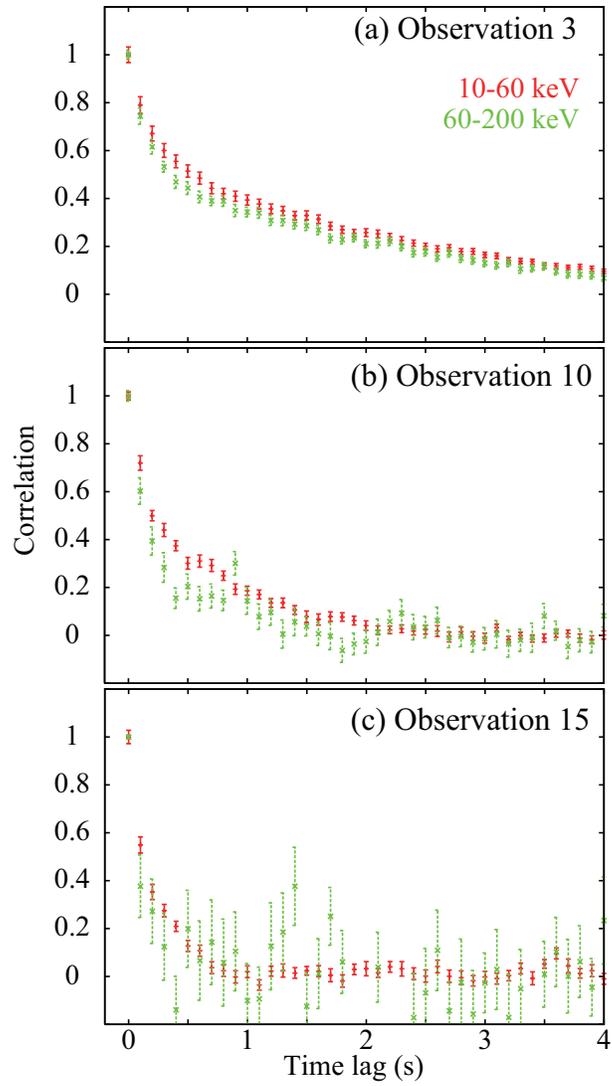}
\end{center}
\caption{The PIN (red) and the GSO (green) ACFs of the three representative observations, calculated in the 10--60 keV and 60--200 keV range, respectively.
(a) Observation 3, (b) Observation 10, and (c) Observation 15.}
\label{fig:3ACF}
\end{figure}

Figure \ref{fig:3ACF} displays the ACFs calculated with the PIN and the GSO data for the three representative observations.
Thus, the ACFs monotonically decrease on typical time-lag scales of 0.2--1 s, which become shorter as $C_{\rm{ASM}}$ get higher.
This behavior is in qualitative agreement with that of $\nu_{\rm{b}}$, and the time-lag scale is consistent with the corresponding (2$\pi$$\nu_{\rm{b}}$)$^{-1}$.

\begin{figure}
\begin{center}
\FigureFile(80mm,75mm){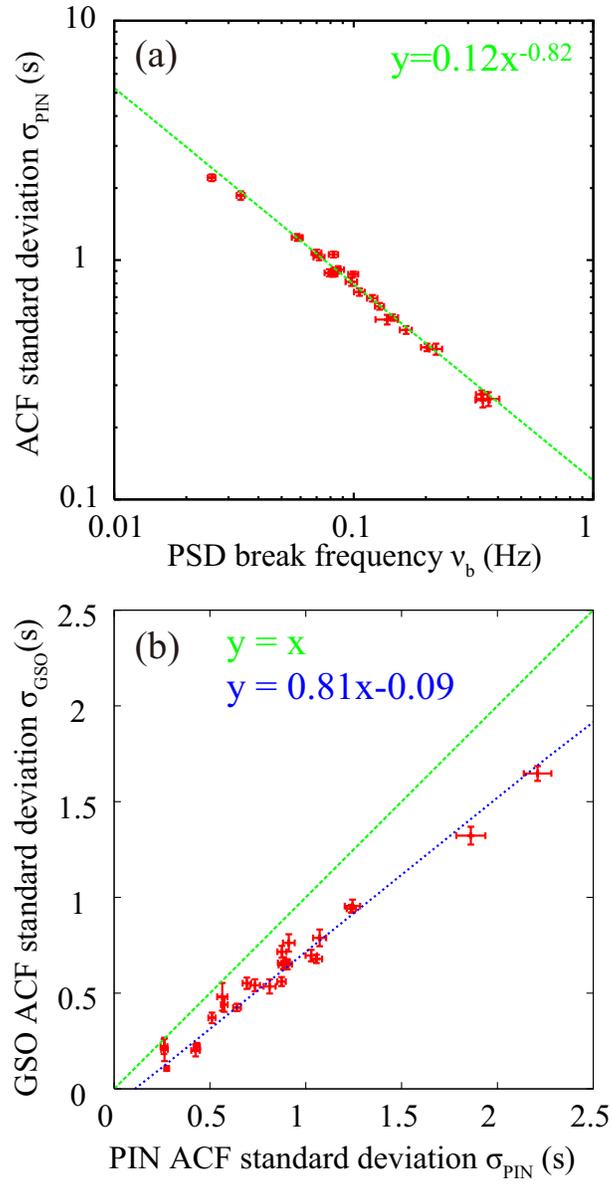}
\end{center}
\caption{(a) The characteristic time scale of the 10--60 keV PIN ACF $\sigma_{\rm{PIN}}$, shown against $\nu_{\rm{b}}$. (b) Correlation between $\sigma_{\rm{PIN}}$ and $\sigma_{\rm{GSO}}$.}
\label{fig:ACFvsPSD}
\end{figure}

We tried quantifying these ACFs by calculating their characteristic time scales using their standard deviation from the zero lag, namely, 
\begin{eqnarray}
\sigma=\sqrt{\frac{\sum^{}_{i} ACF(t_i) * {t_i}^2}{\sum^{}_{i} ACF(t_i)}}
\label{disp}
\end{eqnarray}
where $t_{\rm{i}}$ is the time lag at the i-th bin. Hereafter, we denote $\sigma$ for the PIN ACF as ${\sigma}_{\rm{PIN}}$, and that for the GSO ACF as ${\sigma}_{\rm{GSO}}$.
For the PIN data, the summation was taken from zero lag to the point where correlation value first reaches 0.01.
In calculating ${\sigma}_{\rm{GSO}}$, the summation was taken over the same time lag range as the corresponding PIN data.
Figure \ref{fig:ACFvsPSD} (a) shows the obtained $\sigma_{\rm{PIN}}$ against $\nu_{\rm{b}}$.
The tight anti-correlation between the two quantities confirms the expected consistency between the PSDs and the ACFs.

Since an ACF is generally less subject to the Poisson noise than the corresponding PSD, the GSO ACFs, such as shown in figure \ref{fig:3ACF} (green), allow us to study the reported energy dependence of the Cyg X-1 variability (Miyamoto et al. 1989; Maccarone et al. 2000). Actually, figure \ref{fig:3ACF} suggests that $\sigma_{\rm{GSO}}$ is somewhat smaller than $\sigma_{\rm{PIN}}$.
In order to examine this possibility, $\sigma_{\rm{PIN}}$ is plotted against $\sigma_{\rm{GSO}}$ in figure \ref{fig:ACFvsPSD} (b). 
As suggested by the three representatives, $\sigma_{\rm{GSO}}$ is thus confirmed to be always smaller than $\sigma_{\rm{PIN}}$ by $\sim$ 20 \% in our data sets. 

\subsection{Cross Correlation Function}
Previous GINGA observations of Cyg X-1 discovered interesting ``phase-lag" phenomena among variations in different energy bands (Miyamoto et al. 1989).
As a brief examination of such effects in the present data which span a much broader energy band than the Ginga data (typically 2--30 keV), we computed,
using $\tt{crosscor}$ in XRONOS, cross correlation functions (CCFs) between the 10--60 keV PIN data and the 60-200 keV GSO data, with the same
conditions as employed in the ACF analysis. More detailed cross spectrum analysis is a subject of our future work.
Each CCF was normalized to the geometric mean of the underlying two ACFs, from which the Poisson noise contribution was subtracted as in subsection \ref{sec:ACF}.

\begin{figure}
  \begin{center}
\FigureFile(80mm,80mm){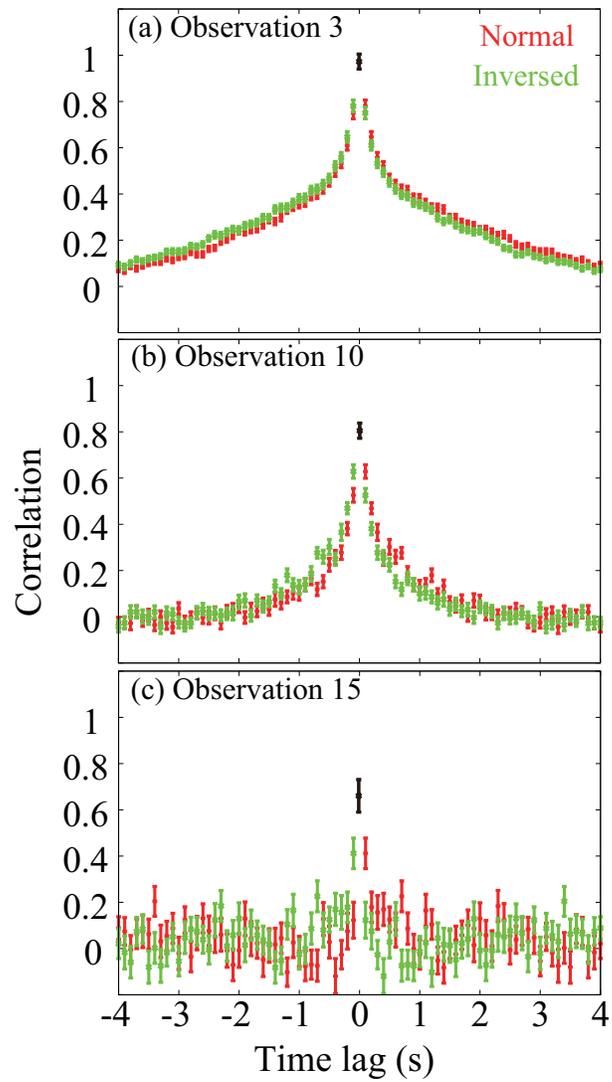}
  \end{center}
  \caption{CCFs between the 10--60 keV PIN data and the 60--200 keV GSO data from the three representative observations (red), with time-inverse ones (green). The points at zero lag are presented with black points.}
  \label{fig:3CCF}
\end{figure}

The obtained CCFs of the three representatives are shown in figure \ref{fig:3CCF}. 
Unlike the ACFs, their peak values become less than 1 if variations in the two energy bands are not perfectly correlated with each other.
Let us examine these CCFs for their lower-to-higher-order moments. As to the first-order property, the three CCFs in figure \ref{fig:3CCF} are all peaked
at zero time lag within the employed time resolution of 0.1 s. This is consistent with previous works in energies below $\sim$ 100 keV,
including an RXTE upper limit of 2 ms on time lags between 2--5 and 24--40 keV signals (Maccarone et al. 2000).

Next, from the second-order view, we confirm in figure \ref{fig:3ACF} that the characteristic time scale of CCFs gets significantly shorter toward higher values of $C_{\rm{ASM}}$, as readily expected
from the ACF analysis.
What becomes possible with CCFs is asymmetry against the time reversal, namely the third-order moment, which is related to phase-lag properties between the two energy bands.
In fact, we find in figure \ref{fig:3CCF} that the characteristic time scale of correlation toward the positive time lag is longer than toward the negative time lag.
This means that higher energy photons
have some delayed components against softer ones, in qualitative argument with the reported hard-phase-lags (Miyamoto et al. 1989; Nowak et al. 1999; Maccarone et al. 2000; Kotov et al. 2001; Pottschmidt et al 2003; Ar\'{e}valo \& Uttley 2005).

\begin{figure}
  \begin{center}
\FigureFile(80mm,80mm){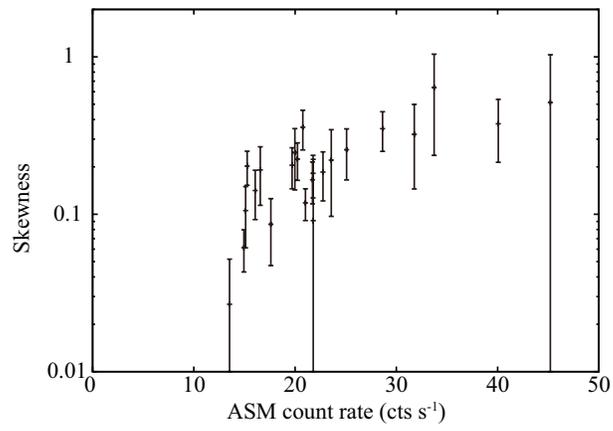}
  \end{center}
  \caption{Dependence of the CCF skewness on $C_{\rm{ASM}}$. The data point for Observation 16 ($C_{\rm{ASM}}$ = 40.2 cts s$^{-1}$) is omitted, because of too large errors.}
  \label{fig:CCFasym}
\end{figure}

To study this property more quantitatively, we calculated skewness as,
\begin{eqnarray}
S = \frac{\sum^{}_{i} CCF(t_i) * (\bar{t}-t_i)^{3}}{ \sum^{}_{i} CCF(t_i)*[\sum^{}_{i}CCF(t_i)*(\bar{t}-t_i)^{2}]^{\frac{3}{2}}}
\label{area}
\end{eqnarray}
where $\bar{t}$ is the centroid of the CCF distribution. In every observation, $\bar{t}$ was calculated as $\sim$ 0.1 s or smaller.
The range employed for this summation is the standard deviation from the center of each CCF since the calculated skewness, the third-order quantity of time,
could have too large uncertainties if we continued the calculation until zero correlation. By the definition of our CCF time lag, $S$ $>$ 0 means hard-phase lag.
The results of these calculations, given in figure \ref{fig:CCFasym}, reveal that the skewness of every CCF is larger than 0, so there are hard-lags in all observations.

\section{Discussion}

We have analyzed the Suzaku HXD data of Cyg X-1,
acquired on 25 occasions from 2005 September through 2009 December,
and studied the 10--400 keV spectral properties
as well as the 10$^{-3}$--10 Hz timing behavior.
The source resided in the LHS throughout,
accompanied by  factor $\sim 3$ changes
in the 1.5--12 keV RXTE ASM intensity, $C_{\rm{ASM}}$,
which was used as an indicator of the soft X-ray flux and of the mass accretion rate.
Among these data sets, the first one was already analyzed in Paper I.
Thanks to the high performance of the HXD including its GSO scintillators,
the present study has provided considerably improved
hard X-ray information on this leading BHB,
particularly in energies above $\sim 100$ keV,
compared with previous studies
(e.g. Miyamoto et al. 1989; Gierli\'{n}ski et al. 1997; Gilfanov et al. 1999; Nowak et al. 1999;
Maccarone et al. 2000; Di Salvo et al. 2001; Frontera et al. 2001a; Pottschmidt et al. 2003;
Zdziarski \& Gierli\'{n}ski 2004; Ibragimov et al. 2005; Titarchuk et al. 2007; Gierli\'{n}ski et al. 2010).

\subsection{Results and interpretation from spectral analyses}
\label{discuss:subsec:spectra}

The 25 HXD spectra covering 10--400 keV were all reproduced successfully
by a single {\tt compPS} model,
incorporating reflection from a cold matter.
Therefore, as envisaged repeatedly in the literature (section 1)
including Paper I in particular,
the hard X-ray production in Cyg X-1 can be interpreted
in the disk-corona framework;
a disk provides  soft X-ray seed photons,
which are  Comptonized by a hot corona,
and the produced hard X-ray photons are
partially reflected by the disk.

In terms of  these {\tt compPS} fits,
the 10--400 keV energy  flux in the 25 observations was calculated  as
(2--5) $\times 10^{-8}$ erg cm$^{-2}$ s$^{-1}$ [figure \ref{fig:specpara} (a)].
Through very crude analysis of some representative  of the XIS data,
we estimated the 0.1--10 keV  flux in our observations
as  (2--4) $\times 10^{-8}$ erg cm$^{-2}$ s$^{-1}$,
after removing photoelectric absorption.
Therefore,  the  unabsorbed flux in the 0.1--400 keV range,
where most of the emission is contained (cf. figure 10 in Paper I),
is estimated as (4--8) $\times 10^{-8}$ erg cm$^{-2}$ s$^{-1}$ in our observations.
Assuming isotropic emission and a distance of 2 kpc (Zi{\'o}{\l}kowski 2005),
these values correspond to 0.8--1.6\% of the  Eddington luminosity
for a  black hole of 16 solar masses (Caballero-Nieves et al. 2009).
It would be interesting to compare the state of Cyg X-1 between our data sets
and a large amount of pointed observations conducted by RXTE (Gierli\'{n}ski et al. 2010).
In terms of the ``state parameter" defined in Gierli\'{n}ski et al. (2010),
the 25 Suzaku data are found in the range of 0--0.25,
which is very similar to the LHS covered by the RXTE observations.
The present results thus provide unified broad-band information on this leading BHB over a typical,
and reasonably wide, flux range in the LHS. 

The wide-band spectroscopy with Suzaku has successfully decoupled
the intrinsic thermal cutoff in the continuum and the reflection hump,
and gave the reflection solid angle $\Omega/2\pi$ as 0.20--0.35.
This parameters was previously obtained by
Gilfanov et al. (1999) as 0.4--0.6 with RXTE, by Ibragimov et al. (2005) as 0.2--0.5 with Ginga, OSSE and RXTE,
while as 0.1--0.2 by Gierli\'{n}ski et al. (2010) with RXTE.
Thus, the absolute values can vary depending on the model and the band width utilized for analysis.
The cool disk is therefore thought to penetrate halfway into the corona (Paper I),
because these values of $\Omega/2\pi$ would be too small 
if the disk reached the innermost stable circular orbit,
while too large if the disk were detached from the corona.

The {\tt compPS} fits have revealed  some interesting dependences
of the Compton parameters on $C_{\rm{ASM}}$ (figure~\ref{fig:specpara}).
Although neither $T_{\rm{e}}$ nor $\tau$ exhibits noticeable systematic behaviors,
their product, $y$, clearly decreases as $C_{\rm{ASM}}$ increases [figure~\ref{fig:specpara}(b)].
Therefore, the effects of Comptonization on individual photons
are  considered to diminish as the soft X-ray flux becomes higher.
Then, the positive dependence of the HXD flux on $C_{\rm{ASM}}$  [figure~\ref{fig:specpara}(a)]
must be a result of higher seed photon inputs to the Compton corona
towards higher mass accretion rates. Further considering the positive $\Omega/2\pi$ vs $C_{\rm{ASM}}$ 
correlations as noted above,the disk-corona geometry must evolve towards higher accretion rates
in such a way that the energy transfer from the corona to the disk photons decreases,
while the seed photon input and the disk reflection both increase.
After discussing the time variations,
we come back to this issue in subsection 5.3.

Our single-zone modeling is admittedly  rather approximate,
because the corona, in reality, is likely to be inhomogeneous;
a full-band spectral modeling including the XIS data would require
multiple Compton components with different  $y$-parameters (Paper I).
Actually, our  results on Observation 1 are somewhat discrepant
from those of the harder Compton component
which comprises the double Comptonization model  employed in Paper I.
For example, our $y$-parameter, $0.87$ (table~2),
is smaller (softer) than that of the harder Compton component in Paper I,
$y=1.15^{+0.05}_{-0.03}$,
because in that modeling some low-energy signals
were taken up by the  softer Compton component.
The reflection solid angle we obtained ($\Omega/2\pi=0.20$; table~2) 
could also be smaller than that of Paper I ($\Omega/2\pi=0.4^{+0.2}_{-0.3}$),
probably because a slightly concave  double Compton continuum
requires a stronger reflection to compensate  for it.
Nevertheless, we consider our single-zone approach to be  meaningful,
because the single {\tt compPS} parameters  derived here from the HXD data are
regarded as some kind of averages over the suggested coronal inhomogeneities.

Incidentally, we obtained  somewhat lower values of  $T_{\rm{e}}$ than in Paper I.
This effect is  mostly due to  the updated GSO data reprocessing (Yamada et al. 2011, in press),
rather than to the modeling difference.

\subsection{Results and interpretation from timing analysis}
\label{discuss:subsec:timing}

The HXD onboard Suzaku, with a relatively small effective area
(e.g., $\sim 270$ cm$^{2}$ at 100 keV; figure~10 of Takahashi et al. 2007),
is not necessarily optimized for studies of fast aperiodic variations.
Nevertheless, its extremely low background and wide energy band have
allowed us to obtain some improved information on the fast flickering of Cyg X-1,
particularly in energies above $\sim 100$ keV
where previous hard X-ray instruments were
limited by background.

Our timing analysis gave 10$^{-3}$--10 Hz PSDs (figure~\ref{fig:PSD})
which are typical of this object;
a relatively flat power  distribution and a red-noise behavior,
below and above a rather well defined break frequency  $\nu_{\rm b}$, respectively.
The values of $\nu_{\rm b}$  we found,
0.03--0.3 Hz [figure \ref{fig:PSDresult} (b)],
are consistent with the range of 0.03--0.5 Hz
which Pottschmidt et al. (2003) derived from the 2--13 keV RXTE PCA data
wherein $C_{\rm{ASM}}$ varied over 15--80 cts s$^{-1}$.
Our ACF studies have been consistent
with the PSD analyses [figure~\ref{fig:3ACF}, figure~\ref{fig:ACFvsPSD} (a)].

Figure \ref{fig:rmsspec} shows that 
the total fractional rms variations,  integrated over 10$^{-3}$--10 Hz,
were found in the range of 20--40\%, 
which is typical of this object.
In addition, the fractional rms variation was found to be
approximately energy independent up to 100--200 keV.
This reconfirms our basic interpretation  (section 1)
that the X-ray flickering reflects variations
in the seed photon input to the corona. 

As already mentioned in subsection 4.1,
one important result from our timing studies is
the prominent evolution of the PSDs (figure~\ref{fig:PSD}, figure~\ref{fig:PSDresult}):
towards higher values of  $C_{\rm{ASM}}$,
$\nu_{\rm b}$ increases and the low-frequency power decreases,
while the PSD above $\nu_{\rm b}$ is kept rather unchanged.
The change in $\nu_{\rm b}$ is also supported by the ACF behavior in figure~\ref{fig:3ACF}.
Then, without invoking  particular models of variability,
we may argue in the following way.
The observed aperiodic flux changes may be regarded
as a superposition of some local (or propagating) fluctuations,
occurring in the corona (or in the coronal covering fraction over the cool disk)
which is thought to span a broad radius range of
a few to $\sim 100$ gravitational radii (Paper~I).
Furthermore, it is plausible to think that slower (or faster) variations
are produced mainly in outer (or inner) regions of the corona,
where various physical phenomena
generally have longer (or shorter) time scales.
Then, one of the most natural ideas would be to presume
that an increase in the mass accretion rate causes
the outer coronal boundary  to shrink inwards,
and hence  the slowest variations produced in the outermost region disappear.
This can consistently explain the observed three effects;
the increase in $\nu_{\rm b}$, the decrease in low-frequency power,
and the unchanged high-frequency power.

To be somewhat more specific, let us assume, for example,
that the variation time scale is related to viscous time scales
in an standard ``alpha" disk (Shakura \& Sunyaev 1973),
which is described as
\begin{equation}
t_{\rm vis}  \propto \alpha^{-1} (r/H)^{2} {\Omega_{\rm{K}}}^{-1}~.
\label{eq:visc}
\end{equation}
Here,  $\alpha$ is the viscosity parameter, $r$ is the radius,
$H$ is the scale height of the accretion disk or the corona,
and $\Omega_{{\rm{K}}}$ is the Keplerian angular velocity.
Supposing that $\alpha$ and $r/H$ are both kept unchanged,
then  coronal size changes by a factor of $\sim 5$ are 
sufficient to explain the observed change in $\nu_{\rm b}$
by an order of magnitude among the present data sets.

\subsection{The inferred accretion geometry}

We may now combine the spectral view developed
in  subsection~\ref{discuss:subsec:spectra}
with the variability consideration in subsection~\ref{discuss:subsec:timing}.
In the former subsection, we observed the three spectral evolutionary changes as $C_{\rm{ASM}}$ increased;
the increase in the seed photon input, the decrease in $y$, and the increase in the reflection solid angle.
These results were interpreted in terms of a view developed in Paper I, namely, partial overlap between
the disk and the corona. In the latter subsection, an increase in the mass accretion rate was found to cause three effects, 
namely, the increase in $\nu_{\rm b}$, the increase in the low-frequnecy power and the rather unchanged high-frequency power density. We interpreted these results
that the outer boundary of the corona, to be denoted $R_{\rm c}$,
shrank (e.g., by a factor of 5 from the $\nu_{\rm{b}}$ change) as the mass accretion rate increased.

When $R_{\rm c}$ decreases towards higher accretion rates,
the coronal height $H$ is expected to also reduce in rough proportion to $R_{\rm c}$,
because  the corona is considered to be approximately spherical.
As $y$ is approximately proportional to $n_{\rm e} T_{\rm e} H$
(with $n_{\rm e}$ the average electron density),
the observed concurrent decrease in $y$ can be explained
if the coronal electron pressure $n_{\rm e} k T_{\rm e}$ 
somehow remains approximately constant,
or if it even decreases, e.g., under increased Compton cooling.

Let us consider the behavior of the innermost disk radius $R_{\rm in}$,
in order to explain the remaining two spectral evolutions,
One of them is a need for a larger number of seed photons
to be supplied to the corona at higher accretion rates.
This clearly implies a larger overlap between the corona and disk,
thus requiring $R_{\rm in}$ to decrease by a larger factor than  $R_{\rm c}$.
This view is based on a conclusion of Paper I that the cool disk is 
truncated at $\sim$ 15 $R\rm{_G}$.
Then, the remaining property, namely the increased reflection,
follows as a natural consequence,
because a larger fraction of the corona becomes
penetrated by the cool disk.
The invoked larger disk-corona overlap will also decrease
$T_{\rm e}$ of the corona through increased Compton cooling,
and will contribute to the required reduction in $y$.
Thus, the observed spectral evolution,
as well as those in the variation time scales,
can be explained consistently and very naturally by assuming
that the disk  intrudes more deeply into the corona
as $R_{\rm c}$ decreases.

\subsection{Implications of the energy dependence of time variability}

Finally, let us discuss implications of the  subtle but
intriguing two energy-dependent  effects in fast variations,
as revealed or reconfirmed
utilizing the genuine broad-band capability of the HXD.
One of them is the relation of $\sigma_{\rm{GSO}}<\sigma_{\rm{PIN}}$,
found  in all observations by the ACF analysis (figure \ref{fig:ACFvsPSD}).
In figure~\ref{fig:PSD}, this effect may be visible as a slightly flatter
slope above $\sim 0.3$ Hz in the GSO PSD than in the PIN  PSD. 
In other words, 
harder photons vary systematically on shorter time scales
as observed previously (e.g., Maccarone et al. 2000; Pottschmidt et al. 2003), 
and the effect persists at different mass accretion rates.

In the present disk-corona picture,
the faster variability in harder photons 
can be explained most naturally,
if the multi-$y$ property revealed in Paper I
takes a form of negative radial gradient in $y$.
Then, harder photons are expected to vary on faster time scales,
because they will be produced more efficiently  
in  regions with larger $y$, namely inner regions, 
where the  variability time scales should be shorter.
Though still speculative at present,
a particularly attractive possibility is
that the outer coronal region overlapping with the disk
has  lower $y$ value(s) under photon cooling,
while the non-overlapping inner region has higher $y$.

The other energy-dependent effect is 
the non-zero skewness seen in figure~\ref{fig:CCFasym},
which is essentially equivalent to the CCF asymmetry:
while variations in the 60--200 keV range exhibit no
direct time lags  within 0.1 s from those in 10--60 keV,
their cross correlation is stronger 
when the harder signals are delayed (figure~\ref{fig:3CCF}).
These results extend, up to $\sim$ 200 keV,
the previously reported ``hard-phase lags", namely, the effect 
that a cross spectrum  between softer and harder signals
exhibits phase differences
that depend much less strongly on the frequency 
than in the case of a constant time lag 
(e.g., Miyamoto et al. 1989; Nowak et al. 1999;  Poutanen 2001; Pottschmidt et al. 2003).
Similar hard-phase  lags have been observed from other BHBs in the LHS
(e.g., Miyamoto et al. 1992; van der Hooft 1999ab; Nowak et al. 1999).
As pointed out in many  previous works (e.g., Maccarone et al. 2000),
this  phenomenon in Cyg X-1 and other BHBs
is difficult  to explain in terms of Compton delays,
since this would predict $\sigma_{\rm{GSO}}>\sigma_{\rm{PIN}}$
in contradiction to the observations,
and would require too large a corona
to produce phase lags on time scales of $\sim 1$ s.

A more reasonable explanation to the hard-phase lags is to assume
that fluctuations responsible for the X-ray variations
propagate inwards through the corona
which has a radial gradient in $y$
as invoked above (Lyubarskii 1997; Done et al. 2007).
Then, harder signals,
produced in regions closer to the black hole,
would be generally delayed from softer ones
to be produced in more outer regions.
Furthermore, the delay time between the two bands
would depend on the frequency of variation under consideration:
the delay time should be shorter for higher-frequency fluctuations,
because they must originate in regions closer to the black hole,
where propagation delays would also be shorter.
Then, we expect the energy dependence to be
of phase-lag type than a constant-time-lag type. 
These ideas were already presented by Nowak et al. (1999),
although their main focus was on the Compton scattering. 

In this way, 
the two energy dependent effects of time variability
have reinforced the multi-$y$ picture derived in Paper I.
While Paper I assumed a condition 
wherein domains with high and low values of $\tau$ coexist at each radius,
it has become more likely that a large-scale gradient in $y$ is also present.

\section{Conclusion}

Analyzing  the Suzaku HXD data of  Cyg X-1 acquired
on 25 observations over 2005 October through 2009 December,
we have obtained the following results.

\begin{enumerate}
\item Cyg X-1 was  in the LHS on these occasions,
accompanied by a factor 3 changes in $C_{\rm{ASM}}$.
The estimated 0.1--400 keV unabsorbed flux  varied over
(4--8) $\times$ 10$^{-8}$ erg cm$^{-2}$ s$^{-1}$.

\item
The 10--400 keV HXD spectra were all reproduced  successfully
with a single Comptonization model, 
considering reflection from a cold matter.

\item As $C_{\rm{ASM}}$ became higher,
the HXD flux and the reflection solid angle increased, 
while the Compton $y$-parameter decreased.

\item 
When $C_{\rm{ASM}}$ increased by a factor of $\sim$ 3,
the PSD break frequency became higher by an order of magnitude,
and the power integrated over 10$^{-3}$--10$^{-2}$ Hz
diminished by a factor of $\sim$ 5. 

\item In the ACF analysis, the 60--200 keV signals
always showed shorter correlation lengths
than the 10--60 keV ones.

\item The CCF revealed skewness so that 
the 60-200 keV signals were delayed from those in the  10-60 keV, 
and the effects became more prominent towards  higher mass accretion rate.

\end{enumerate}

In order to consistently explain items 3 and 4 above, 
we have developed an interpretation
that  a gradual increase in the mass accretion  rate
causes the corona to diminish in size,
while the cool disk to intrude more deeply into the corona. 
Property 5 and 6 listed above may be explained
by considering a radial gradient in the corona,
so that harder photons are produced in
closer vicinity of the black hole.

\section*{Acknowledgement}
The research presented in this paper has been financed by
Grant-in-Aid for JSPS Fellows. PG acknowledges a JAXA
International Top Young Fellowship. We thank the referee,
Dr.A.Zdziarski, for his valuable comments.

\end{document}